\documentclass[pdftex,12pt]{JHEP3}
\usepackage{amsfonts,amsbsy,latexsym,amssymb,amscd,amstext,amsmath}
\usepackage[pdftex]{epsfig}
\pdfoutput=1

\renewcommand{\AmS}{{\protect\the\textfont2A\kern-.1667em\lower.5ex\hbox{M}\kern-.125emS}}
 \def\be{\begin{equation}}
\def\ee{\end{equation}}
\def\bea{\begin{eqnarray}}
\def\eea{\end{eqnarray}}
\def\pd{\partial}
\def\a{\alpha}
\def\b{\beta}
\def\g{\gamma}
\def\d{\delta}
\def\m{\mu}
\def\n{\nu}

\def\l{\lambda}

\def\r{\rho}
\def\s{\sigma}

\def\bi{\begin{itemize}}
\def\ei{\end{itemize}}

\date{ 2011} \preprint{IFT-UAM/CSIC-12-38;FTUAM-12-91}
\title{The weight of matter. } \author{Enrique \'Alvarez  \\  Instituto de F\'{\i}sica Te\'orica
UAM/CSIC and Departamento de F\'{\i}sica Te\'orica \\ Universidad
Aut\'onoma de Madrid, E-28049--Madrid, Spain \\ E-mail: \email{enrique.alvarez@uam.es }}
\abstract{Einstein's traceless 1919 gravitational theory is analyzed from a variational viewpoint. It is shown to be equivalent to a transverse (invariant only under diffeomorphisms that preserve the Lebesgue measure) theory, with an additional Weyl symmetry, in which the gauge is partially fixed so that the metric becomes unimodular. In spite of the fact that this symmetry forbids direct coupling of the potential energy with the gravitational sector, the equivalence principle is recovered in the unimodular gauge owing to Bianchi's identities.
}
\begin{document}

{\vskip 1cm}
\newpage
%%%%%%%%%%%%%%%
\vskip 1cm

%%%%%%%%%%%%%%%%%%%%%%%%%%%%%%%%%%%%%%%%%%%%%%%%%%%%%%%%%%%%%%%%%%%%%%%%%%%%%%%%%%%%%%%%%%%%%%%%%%%%%%%%%%%%%%%%%%%%%%%%%%%%%%%%%%%%%%%%%%%
\vskip 1cm

%%%%%%%%%%%%%%%%%%%%%%%%%%%%%%%%%%%%%%%%%%%%%%%%%%%%%%%%%%%%%%%%%%%%%%%%%%%%%%%%%%%%%%%%%%%%%%%%%%%%%%%%%%%%%%%%%%%%%%%%%%%%%%%

\newpage
\tableofcontents

%\newpage
%%%%%%%%%%%%%%%%%%%%%%%%%%%%%%%%%%%%%%%%%%%%%%%%%%%%%%%%%%%%%%%%%%%%%%%%%%%%%%%%%%%%%%%%%%%%%%%%%%%%%%%%%%%%%%%%%%%%%%%%%%%%%%%%%%%%%%%%%%%%%%%%%%%%%%%%%%%%%%%%%%%%%%%%%%%%%%%%%%%%%%%%%%%%%%%%%%%%%%%%%%%%%%%%%%

\setcounter{page}{1}
\setcounter{footnote}{1}
\tableofcontents

 %%%%%%%%%%%%%%%%%%%%%%%%%%%%%%%%%%%%%%%%%%%%%%%%%%%%%%%%%%%%%%%%%%%%%%%%%%%%%%%%%%%%%%%%%%%%%%%%%%%%%%%%%%%%%%%%%%%%%%%%%%%%%%
\section{Introduction}
%%%%%%%%%%%%%%%%%%%%%%%%%%%%%%%%%%%%%%%%%%%%%%%%%%%%%%%%%%%%%%%%%%%%%%%%%%%%%%%%%%%%%%%%%%%%%%%%%%%%%%%%%%%%%%%%%%%%%%%%%%%%%
Let us consider a quite general theory of interacting matter in ordinary Minkowski space, general enough to include the main aspects (with the exception of chirality, which does not play an important r\^ole in the considerations that follow) of the standard model of elementary particle physics: fermions, gauge fields and scalars

\bea
&&S_{matt}=\int d^n x~ \sum_{ij} \bar{\psi}_i\left(~i\g^\m(\d_{ij}\pd_\m -\l A^a_\m (T^a)_{ij})+y_{ij}\phi-m\d_{ij}\right)\psi_j +\nonumber\\
&&{1\over 2}~\eta^{\m\n}~\pd_\m\phi\pd_\n\phi-V(\phi)-{1\over 4}\sum_a (F^a_{\m\n})^2
\eea
We have denoted the gauge coupling by $\l$ in order to avoid confusion with the determinant of the metric tensor.
To ask the question of how much does the matter weigh is the same as to ask what is the coupling to the gravitational field $g_{\m\n}$. The standard answer (known as minimal covariant coupling) \footnote{The seemingly stablished (although theoretically poorly understood) fact that only color singlet states appear in the physical spectrum means that it is not clear whether minimal coupling should be implemented at the level of the quarks (as we have done here) or else at the level of the physical states, which only have a lagrangian description in a effective theory.
}
posits a theory invariant under the full diffeomorphism group $\text{Diff(M)}$ of the space-time manifold and reads
\bea\label{standardmodel}
&&S_{matt}=\int~\sqrt{|g|} d^n x~ \sum_{ij}~{i\over 2}~e_a^\m\left(\bar{\psi}_i~ \g_a~D_\m\psi_i-D_\m\bar{\psi}_i \g^a \psi_i\right)+\nonumber\\
&&\bar{\psi}_i\left( -i g\g_a e_a^\m  A^b_\m (T^b)_{ij}+y_{ij}\phi-m\d_{ij}\right)\psi_j +
{1\over 2}~g^{\m\n} \pd_\m\phi\pd_\n\phi-V(\phi)-{1\over 4}\sum_a F^a_{\m\n}g^{\m\a}g^{\n\b}F^a_{\a\b}\nonumber
\eea

where the spin-covariant derivative is given by
\[
D_\m \psi\equiv \left(\pd_\m+S_\m^{ab}\Sigma_{ab} \right)\psi
\]
\[
\Sigma_{ab}\equiv {1\over 2}\g_{ab}
\]
\[
S_\m^{ab}\equiv {1\over 2}\left(e^a_\r\nabla_\m e^ b_\s-e^b_\r\nabla_\m e^a_\s\right)g^{\r\s}
\]
 
This leads to the {\em energy-momentum tensor}  representing physically the active gravitational source-density.
\[
\d S_{matt}\equiv {1\over 2 }\int \sqrt{|g|} d^n x T_{\m\n}\d g^{\m\n}(x)
\]
with
\bea
&&T_{\m\n}\equiv \sum_{ij}~{i\over 2}\left(\bar{\psi}_i\g_{(\m}D_{\n)}\psi_i-D_{(\m}\bar{\psi}_i\g_{\n)}\psi_i\right)+\bar{\psi}_i\left( i g A^a_\m (T^a)_{ij}-y_{ij}\phi+m\d_{ij}\right)\psi_j +\nonumber\\
&&\pd_\m\phi\pd_\n\phi-g_{\m\n}\left({1\over 2}g^{\a\b}\pd_\a\phi\pd_\b\phi-V(\phi)\right)+{1\over 4}g_{\m\n}F_{\a\b}F^{\a\b}-F_{\m}\,^\r F_{\r\n}
\eea
 
 This the archetypical energy-momentum tensor, and we shall use this word in this sense only, reserving the name {\em active gravitational source} for the variation of the action with respect to the metric in other contexts to be considered shortly.
 \par
 If we now include the Einstein-Hilbert lagrangian
 \[
 S_{EH}\equiv -{1\over 16\pi G}\int d^n x \sqrt{|g|} R
 \]
 the Euler-Lagrange equations of motion (EM) for the metric read
 \[
 R_{\m\n}-{1\over 2}R~g_{\m\n}=8\pi G~ T_{\m\n}
 \] 
 Einstein's (1915) equations. 
 \par
 It is well known (cf. for example, \cite{Alvarez} where many original references may be found) that Einstein (1919) put forward another theory of gravity, which apparently only determines the traceless piece of Einstein's equations
 \[
 R_{\m\n}-{1\over n}R~g_{\m\n}=8\pi G~\left( T_{\m\n}-{1\over n}T~g_{\m\n}\right)
 \] 
 
 This is of course illusory, because Bianchi's identities lead back to 1915 equations with a cosmological constant
 \[
 {n-2\over 2n}R+{8\pi G\over n }T\equiv-\l
 \]
 (precisely the trace of  Einstein's equations of  ordinary general relativity) this equation can be interpreted as an equation for R in case T is known, or else as a constraint on T in case the value of T is unknown.
 \[
 R_{\m\n}-{1\over 2}\left(R+2\l\right)~g_{\m\n}=8\pi G T_{\m\n}
 \] 
 
 These 1919 equations appear as tensorial, diffeomorphism (Diff) invariant equations. The starting point of this paper is whether it is possible to derive them from a Diff invariant action through a variational principle. Einstein's 1919 equations have been already analyzed from a canonical viewpoint by Unruh \cite{Unruh}.

 %%%%%%%%%%%%%%%%%%%%%%%%%%%%%%%%%%%%%%%%%%%%%%%%%%%%%%%%%%%%%%%%%%%%%%%%%%%%%%%%%%%%%%%%%%%%%%%%%%%%%%%%%%%%%%%%%%%%%%%%%%%%%%
\section{The traceless Fierz-Pauli equations.}
 %%%%%%%%%%%%%%%%%%%%%%%%%%%%%%%%%%%%%%%%%%%%%%%%%%%%%%%%%%%%%%%%%%%%%%%%%%%%%%%%%%%%%%%%%%%%%%%%%%%%%%%%%%%%%%%%%%%%%%%%%%%%%%
%%%%%%%%%%%%%%%%%%%%%%%%%%%%%%%%%%%%%%%%%%%%%%%%%%%%%%%%%%%%%%%%%%%%%%%%%%%%%%%%%%%%%%%%%%%%%%%%%%%%%%%%%%%%%%%%%%%%%%%%%%%%%%%
Let us start by examining the problem at the linearized level. The quadratic action in flat space called WTDiff in \cite{AlvarezBGV}\cite{AlvarezV}
 reads
 \[
 L={1\over 4}\pd_\m h^{\n\r}\pd^\m h_{\n\r}-{1\over 2}\pd_\m h^{\m\r}\pd_\n h^\n_\r+{1\over n}\pd^\m h \pd^\r h_{\m\r}-{n+2\over 4 n^2}\pd^\m h \pd_\m h
 \]

Besides being invariant under the transverse Fierz-Pauli (TFP) \footnote{This gauge symmetry is precisely the smallest gauge symmetry necessary in order for the massless limit of a massive theory to exist, even at the linearized level, that is, to reduce the helicity states from 5 in the massive theory down to 2 in the massless one \cite{Bij}} abelian gauge  symmetry, 
\[
\d h_{\m\n}=\pd_\m \xi_\n+\pd_\n \xi_\m
\]
where the parameter $\xi(x)$ is {\em transverse},
\[
\pd_\r \xi^\r=0,
\]
 this lagrangian has got a sort of Weyl invariance
\[
\d h_{\m\n}=\phi(x)\eta_{\m\n}
\]
It is work remarking that the trace h is invariant under TFP:
\[
\d_{\xi_T}h=0
\]
but under Weyl-like
\[
\d_\phi h=n\phi(x)
\]
and denoting 
\[
V_\m\equiv \pd^\r h_{\m\r}
\]
\[
\d V_\m=\Box \xi_\m+\pd_\m\phi
\]
so that
\[
\d \pd.V=\Box \phi
\]
and it is generically possible to choose the gauge
\[
V_\m=f_\m (h)
\]
The corresponding equations of motion read
\[
TFP\equiv~\Box h_{\m\n}-\pd_\m\pd^\r h_{\r\n}-\pd_\n\pd^\r h_{\r\m}-{n+2\over n^2}\eta_{\m\n}\Box h+{2\over n}\left(\pd_\m\pd_\n h+\eta_{\m\n} \pd^\a\pd^\b h_{\a\b}\right)=0
\]
{\em id est}, the traceless piece of the full Fierz-Pauli EM 
\footnote{It is well-known that the FP EM are transverse; acting with $\pd_\m$ on the Fierz-Pauli EM gives the identity
\[
\Box V_\n-\pd_\n\Box h-\Box V_\n-\pd_\n \pd.V+\Box \pd_\n h+\pd_\n \pd.V\equiv 0
\]
On the other hand, the trace yields
\[
(n-2)\left(\pd.V-\Box h\right)=0
\]
so that as long as our gauge condition implies
\[
V_\m={1\over 2}\pd_\m h
\]
(which is always possible) then also
\[
\Box h=0
\]
and the FP EM imply the wave equation for the tensor field:
\[
\Box h_{\m\n}=0
\]
}

\[
FP\equiv~\Box h_{\m\n}-\eta_{\m\n}\Box h-\pd_\m\pd^\r h_{\r\n}-\pd_\n\pd^\r h_{\r\m}+\pd_\m\pd_\n h+\eta_{\m\n} \pd^\a\pd^\b h_{\a\b}=0
\]

Coming back to TFP, the trace vanishes identically, but taking $\pd_\m$ on TFP yields
\[\label{t}
{n-2\over n}\pd_\n\left({1\over n} \Box h-\pd.V\right)=0
\]
Following what it is usually done in FP, choosing a gauge
\[
V_\m={1\over n} \pd_\m h
\]
with cancel all terms on $\pd_\m V_\n+\pd_\n V_\m$ (but we do not get any further information from [\ref{t}]). The TFT EM then reads
\[
\Box \left(h_{\m\n}-{1\over n}h \eta_{\m\n}\right)=0
\]
that is, d'Alembert's equation for the traceless piece of the tensor field.
It is always possible to partially fix the gauge ny choosing a Weyl transformation such that
\[
h=0
\]
The equation is then the wave equation for the tensor field; with the constraint as above and with transverse symmetry only.
\par
 It does not appear feasible to obtain the traceless Fierz-Pauli equations from theory invariant under a full group generated by arbitrary vectors $\xi^\a$; is is necessary to restrict this invariance to the transverse piece $\pd_\a\xi^\a=0$, an also to impose Weyl invariance, so that the dimension of the symmetry group remains unaltered.

%%%%%%%%%%%%%%%%%%%%%%%%%%%%%%%%%%%%%%%%%%%%%%%%%%%%%%%%%%%%%%%%%%%%%%%%%%%%%%%%%%%%%%%%%%%%%%%%%%%%%%%%%%%%%%%%%%%%%%%%%%%%%%
\section{The non linear 1919 variational principle}
%%%%%%%%%%%%%%%%%%%%%%%%%%%%%%%%%%%%%%%%%%%%%%%%%%%%%%%%%%%%%%%%%%%%%%%%%%%%%%%%%%%%%%%%%%%%%%%%%%%%%%%%%%%%%%%%%%%%%%%%%%%%%%%
Let is first consider the vacuum equations. The first thing that comes to mind in order to recover 1919 as an EM is to   to modify the Einstein-Hilbert lagrangian  using as variable
\[
\hat{g}_{\m\n}\equiv~g^{-{1\over n}}g_{\m\n}
\]
which is unimodular in the sense that
\[
\hat{g}=1
\]
First of all, this is not an invariant definition, because after a diffeomorphism with jacobian $j\equiv \text{det}{\pd x^\a~^\prime\over \pd x^\b}$
\[
\hat{g}^\prime_{\m\n}\equiv g^\prime~^{-{1\over n}}~g^\prime_{\m\n}\neq {\pd x^\a\over \pd x^\m~^\prime}{\pd x^\b\over \pd x^\n~^\prime}\hat{g}_{\a\b}
\]
owing to the fact that
\[
g^\prime=j^{-2} g
\]
This means that the corresponding theory is invariant only under the subgroup of $Diff(M)$ that leaves invariant the Lebesgue measure (that is, that get unit jacobian, $j=1$), subgroup the we dubbed TDiff ({\em transverse diffeomorphisms}). Their infinitesimal generator obeys
\[
\pd_\a \xi^\a=0
\]

The theory however is invariant under arbitrary Weyl transformations
\[
g_{\a\b}\rightarrow e^{\s(x)}~g_{\a\b}
\]
The total symmetry is then the combination of both, let us call it WTdiff.
\par
Using the well-known formulas for Weyl transformations, the unimodular action reads
\[
S\equiv\int d^n x~\hat{R}=\int d^n x~g^{1\over n}~\left(R+{n-1\over n}{\nabla^2 g\over g}-{(n-1)(5n-2)\over 4 n^2}{(\nabla g)^2 \over g^2}\right)
\]
Integration by parts yields
\[
S=\int d^n x~\hat{R}=\int d^n x~g^{1\over n}\left(R+{(n-1)(n-2)\over 4 n^2}{(\nabla g)^2 \over g^2}\right)
\]

and using the results of \cite{AlvarezV} the equations of motion read
\bea
&&{\d S\over \d g^{\a\b}}=|g|^{1\over n}\left({1\over n} R g_{\a\b} - R_{\a\b} + {(4 n+3)(n-2)\over 4 n^2} g^{-2} \nabla_\a g \nabla_\b g+{ n -2\over 2n} g^{-1} \nabla_\a \nabla_\b g - \right.\nonumber\\
&&\left.- g_{\a\b}\left({(2n-3)(n-1)(n-2)\over 2 n^2}~g^{-2}~g^{\m\n}\pd_\m g\pd_\n g+{n-2\over 2 n^2}g^{-1}\pd_\m\left(g^{\m\n}\pd_\n g\right)\right)\right)
\eea

\par
On the other hand, it is clear that we can use the gauge
\[
g(x)=F(x)
\]
where $F(x)$ is an arbitrary but fixed function. In the gauge $g=1$, in particular, Einstein's 1919 
equations in vacuum are recovered. Once this gauge freedom is fixed, all that remains is transverse gauge symmetry.

%%%% %%%%%%%%%%%%%%%%%%%%%%%%%%%%%%%%%%%%%%%%%%%%%%%%%%%%%%%%%%%%%%%%%%%%%%%%%%%%%%%%%%%%%%%%%%%%%%%%%%%%%%%%%%%%%%%%%%%%%%%%%%%%%%
 \section{The coupling to matter and the equivalence principle.}
 %%%%%%%%%%%%%%%%%%%%%%%%%%%%%%%%%%%%%%%%%%%%%%%%%%%%%%%%%%%%%%%%%%%%%%%%%%%%%%%%%%%%%%%%%%%%%%%%%%%%%%%%%%%%%%%%%%%%%%%%%%%%%%
Let us assume that the coupling to matter proceeds in such a way as to get local Weyl invariance 
\[
g^\prime_{\m\n}\equiv e^{2\s(x)}g_{\m\n}
\]
 This has been explored in \cite{AlvarezV} and assuming inert matter fields the allowed terms are 

\bea\label{wstandardmodel}
&&S^\text{dirac}=\int~ d^n x~ \sum_{ij}~{i\over 2}~|g|^{1\over 2 n}~e_a^\m\left(\bar{\psi}_i~ \g_a~D_\m\psi_i-D_\m\bar{\psi}_i \g^a \psi_i\right)\nonumber\\
&&S^\text{gauge/dirac}\equiv \int d^n x~-i \l |g|^{1\over 2 n}\sum_{ij}\bar{\psi}_i ~\g^a e_a^\m A^b_\m (T^b)_{ij}\psi_i\nonumber\\
&&S^\text{yukawa}\equiv \int d^n x~\sum_{ij}\left(\bar{\psi}_i y_{ij}\phi-m\d_{ij}\right)\psi_j \nonumber\\
&&S^\text{scalar}\equiv \int d^n x~\left({1\over 2}~|g|^{1\over  n}~g^{\m\n} \pd_\m\phi\pd_\n\phi-V(\phi)\right)\nonumber\\
&&S^\text{gauge}\equiv \int d^n x~-{1\over 4}|g|^{2\over  n}\sum_a F^a_{\m\n}g^{\m\a}g^{\n\b}F^a_{\a\b}
\eea

The powers of the determinant of the metric are precisely determined by our driving hypothesis of rigid Weyl invariance.

The Weyl symmetry we have imposed is  a pure gravitational one; all matter fields are inert under it. The source of the gravitational field (the {\em active gravitational source}) is then (following the action line by line)
\bea
&&{\d S^\text{dirac}\over \d g^{\m\n}}={i\over 2}|g|^{1\over 2n}\sum_{ij}\left[{1\over 4}\left(e^a_\m\left(\bar{\psi}_i \g_a D_\n \psi_i-D_\n \bar{\psi}_i\g^a \psi_i\right)+e^a_\n\left(\bar{\psi}_i \g_a D_\m \psi_i-D_\m \bar{\psi}_i\g^a \psi_i\right)\right)-\right.\nonumber\\
&&\left.{1\over 2n}e_a^\r\left(\bar{\psi}_i\g^a D_\r\psi_i-D_\r\bar{\psi}_i \g^a\psi_i\right)~g_{\m\n}\right]\nonumber\\
&&{\d S^\text{gauge/dirac}\over \d g^{\m\n}}=-i \l |g|^{1\over 2 n}\sum_{ij}\left[{1\over 4}\left(e^a_\m\bar{\psi}_i\g_a A^b_\n (T^b)_{ij}\psi_j+e^a_\n\bar{\psi}_i\g_a A^b_\m (T^b)_{ij}\psi_j\right)-\right.\nonumber\\
&&\left.{1\over 2n}\bar{\psi}_i ~\g^a e_a^\m A^b_\m (T^b)_{ij}\psi_i g_{\m\n}\right]\nonumber\\
&&{\d S^\text{yukawa}\over \d g^{\m\n}}=0\nonumber\\
&&{\d S^\text{scalar}\over \d g^{\m\n}}{1\over 2} g^{1\over n}\left(\pd_\m\phi\pd_\n\phi-{1\over n}(\pd_\a\phi)^2~g_{\m\n}\right)\nonumber\\
&&{\d S^\text{gauge}\over \d g^{\m\n}}=-{1\over 2} |g|^{2\over n}\left(F^a_{\m\r}F^a_{\n\s} g^{\r\s}-{1\over n} F^a_{\a\b}T_a^{\a\b} g_{\m\n}\right)
\eea
In all cases the result is proportional to the the traceless energy-momentum tensor 
\[
T_{\m\n}-{1\over n}T~g_{\m\n}
\]
multiplied by different powers of $|g|$. In the gauge $|g|=1$ we then recover the full Einstein's 1919 equations with  transverse TDiff symmetry. 
\par
It should be stressed that all reference to the potential energy has disappeared from the source term. The full set of equations just implements a variational principle for Einstein's 1919 \cite{Einstein} equations. It is remarkable that the equivalence principle is, in a sense, recovered in spite of the fact that the coupling to gravitation is different for each different type of matter field (cf. \cite{Damour})

 %%%%%%%%%%%%%%%%%%%%%%%%%%%%%%%%%%%%%%%%%%%%%%%%%%%%%%%%%%%%%%%%%%%%%%%%%%%%%%%%%%%%%%%%%%%%%%%%%%%%%%%%%%%%%%%%%%%%%%%%%%%%%%
 \section{General Relativity in the unimodular gauge.}
 %%%%%%%%%%%%%%%%%%%%%%%%%%%%%%%%%%%%%%%%%%%%%%%%%%%%%%%%%%%%%%%%%%%%%%%%%%%%%%%%%%%%%%%%%%%%%%%%%%%%%%%%%%%%%%%%%%%%%%%%%%%%%%
We can always work in General Relativity (GR) in the unimodular gauge (cf. \cite{Dragon}\cite{Buchmuller}) $\hat{g}=1$.
Einstein's equations (which must be derived before fixing the gauge) then read
\bea
&&R_{\a\b}
-{n-2\over 4 n^2}\left(\left(2n-1\right){\nabla_\a g\nabla_\b g \over g^2}-2n{\nabla_\b\nabla_\a g\over g}\right)-\nonumber\\
&&{1\over 2}\left(R-{(5n-3)(n-2)\over 4 n^2}{\left(\nabla g\right)^2\over g^2}+{n-2\over n}{\nabla^2 g\over g}\right)g_{\a\b}={8\pi G\over c^4}T_{\a\b}[g^{-1 /n}g_{\m\n}]
\eea
where
\bea
&&\left(\nabla g\right)^2\equiv g^{\m\n}\nabla_\m g\nabla_\n g\nonumber\\
&& \nabla^2 g\equiv g^{\m\n}\nabla_\m\nabla_\n g
\eea
If now the unimodular gauge is chosen, we recover the standard Einstein's general relativity equations (not the traceless part).
\par
It is to me remarked that GR in the unimodular gauge at the EM level is also a transverse theory; the remaining gate freedom must obey
\[
\d g=2 \pd_\a \xi^\a=0
\]

This suggests a way of analyzing transverse theories: any such can be considered as a partially gauge fixed covariant theory with some spur ion (compensating) fields (\cite{Wilczek}\cite{AlvarezF}).
 %%%%%%%%%%%%%%%%%%%%%%%%%%%%%%%%%%%%%%%%%%%%%%%%%%%%%%%%%%%%%%%%%%%%%%%%%%%%%%%%%%%%%%%%%%%%%%%%%%%%%%%%%%%%%%%%%%%%%%%%%%%%%%
 \section{Conclusions.}
 %%%%%%%%%%%%%%%%%%%%%%%%%%%%%%%%%%%%%%%%%%%%%%%%%%%%%%%%%%%%%%%%%%%%%%%%%%%%%%%%%%%%%%%%%%%%%%%%%%%%%%%%%%%%%%%%%%%%%%%%%%%%%%

Einstein's 1919 traceless gravitational equations 
\[
 R_{\m\n}-{1\over n}R~g_{\m\n}={8\pi G\over c^4}\left( T_{\m\n}-{1\over n}T~g_{\m\n}\right)
 \] 
have been derived from an action principle in a particular (unimodular) gauge. Even before gauge fixing the theory is not full Diff invariant; but only TDiff; that is invariant under transverse diffeomorphisms, i.e., those that are transverse in the sense that their generator obeys
\[
\pd_\a \xi^\a=0
\]
The theory enjoys however another gauge symmetry, namely Weyl invariance
\[
g^\prime_{\m\n}\equiv e^{2 \s(x)}~g_{\m\n}
\]
so that the total symmetry group possesses four generators per space-time point.  In the unimodular gauge all that remains is transverse TDiff symmetry. Contrary to popular belief, it is easy to check that this theory is different, even at the classical level, from General Relativity in the unimodular gauge (although the residual gauge symmetry there is also TDiff).
\par
It can appear as nothing short of miraculous that in spite of the fact that the scalar potential, for example, is absent from the source term of Einstein's 1919 equations, we are able to solve them modulo only an integration constant, which can be interpreted as the cosmological constant, and the answer is exactly the same as in general relativity with the full energy-momentum tensor as a source, including the potential energy. The reason for this lies behind Bianchi identities, that impose strong constraints both in Einstein's 1915  and 1919 theories.
\par
Nevertheless let us stress the  irony  lying behind the fact that a theory that {\em forbids} (cf.\cite{AlvarezV}) any direct coupling of the potential energy with the gravitational field is fully equivalent in the gravitational sector (modulo the slightly different symmetry group)  to another one that does allow this coupling, namely general relativity with a cosmological constant. The equivalence principle is besides preserved \cite{Buchmuller}\cite{Damour} in a quite nontrivial way.
\par
It is interesting to check the behavior of this theory at the one loop level. Some indirect computations have already been done in \cite{Kreuzer}\cite{AlvarezFV}, but the setup here is slightly different. We hope to be able to report on work on this topic in the future.

%%%%%%%%%%%%%%%%%%%%%%%%%%%%%%%%%%%%%%%%%%%%%%%%%%%%%%%%%%%%%%%%%%%%%%%%%%%%%%%%%%%%%%%%%%%%%%%%%%%%%%%%%%%%%%%%%%%%%%%%%%%%%%%%
\section*{Acknowledgments}
%%%%%%%%%%%%%%%%%%%%%%%%%%%%%%%%%%%%%%%%%%%%%%%%%%%%%%%%%%%%%%%%%%%%%%%%%%%%%%%%%%%%%%%%%%%%%%%%%%%%%%%%%%%%%%%%%%%%%%%%%%%%%
This work stems from discussions that go back some years with my former collaborators Diego Blas, Anton Faedo, Jaume Garriga, Enric Verdaguer, Roberto Vidal and Juanjo Villarejo. I am grateful to them all.
This work has been partially supported by the European Commission (HPRN-CT-200-00148) as well as by FPA2009-09017 (DGI del MCyT, Spain) and S2009ESP-1473 (CA Madrid).

%%
%%%%%%%%%%%%%%%%%%%%%%%%%%%%%%%%%%%%%%%%%%%%%%%%%%%%%%%%%%%%%%%%%%%%%%%%%%%%%%%%%%%%%%%%%%%%%%%%%%%%%%%%%%%%%%%%%%%%%%%%%%%%%%%%%

\end{document}